\documentclass[aps,prb,amsmath,amssymb,twocolumn,showpacs,superscriptaddress,10pt]{revtex4-2}

\usepackage[english]{babel}
\usepackage{amsmath,amssymb,amsfonts}
\usepackage{graphicx}
\usepackage[colorlinks=True,linkcolor=blue,citecolor=blue,urlcolor=blue]{hyperref}
\usepackage{bookmark}
\usepackage[dvipsnames]{xcolor}
\usepackage{braket}
\usepackage{nicefrac}
\usepackage{bm}
\usepackage{bbm}
\usepackage{slashed}
\usepackage[normalem]{ulem}
\usepackage{array}
\usepackage{makecell}
\newcolumntype{C}{>{$}c<{$}}
\usepackage{lipsum}
\usepackage{nicematrix}
\usepackage{comment}
\usepackage{multirow}

\def\bk{\mathbf{k}}

\newcommand{\tr}{\mathop{\mathrm{tr}}}

\renewcommand{\Re}{\mathop{\mathrm{Re}}}
\renewcommand{\Im}{\mathop{\mathrm{Im}}}

\allowdisplaybreaks
\DeclareMathAlphabet{\zc}{OT1}{pzc}{m}{it}

  \DeclareUnicodeCharacter{2212}{-}

\usepackage{amsthm}

\begin{document}

\title{The analytically tractable zoo of similarity-induced exceptional structures}

\author{Anton Montag}
\email{anton.montag@mpl.mpg.de}
\affiliation{Max Planck Institute for the Science of Light, Staudtstraße 2, 91058 Erlangen, Germany}
\affiliation{Department of Physics, Friedrich-Alexander Universität Erlangen-Nürnberg, Staudtstraße 7, 91058 Erlangen, Germany}

\author{Jordan Isaacs}
\affiliation{Max Planck Institute for the Science of Light, Staudtstraße 2, 91058 Erlangen, Germany}
\affiliation{University of Ottawa, 75 Laurier Avenue, Ottawa, Ontario K1N 6N5, Canada}

\author{Marcus Stålhammar}
\email{marcus.backlund@physics.uu.se}
\affiliation{Institute for Theoretical Physics, Utrecht University, Princetonplein 5, 3584CC Utrecht, The Netherlands}
\affiliation{Department of Physics and Astronomy, Uppsala University, Uppsala, Sweden}

\author{Flore K. Kunst}
\email{flore.kunst@mpl.mpg.de}
\affiliation{Max Planck Institute for the Science of Light, Staudtstraße 2, 91058 Erlangen, Germany}
\affiliation{Department of Physics, Friedrich-Alexander Universität Erlangen-Nürnberg, Staudtstraße 7, 91058 Erlangen, Germany}

\date{\today}

\begin{abstract}
Exceptional points (EPs) are non-Hermitian spectral degeneracies marking a simultaneous coalescence of eigenvalues and eigenvectors.
Despite the fact that multiband $n$-fold EPs (EP$n$s) generically emerge as special points on manifolds of EP$m$s, where $m<n$, EP$n$s as well as their topological properties have hitherto been studied as isolated objects.
In this work, we address this issue and carefully map out the emerging properties of multifold exceptional structures in three and four dimensions under the influence of one or multiple generalized similarities, revealing diverse combinations of EP$m$s in direct connection to EP$n$s.
We find that simply counting the number of constraints defining the EP$n$s is not sufficient in the presence of similarities; the constraints can also be satisfied by the EP$m$ manifolds obeying certain spectral symmetries in the complex eigenvalue plane, reducing their dimension beyond what is expected from counting the number of constraints.
Furthermore, the induced spectral symmetries not always allow for any EP$m$ manifold to emerge in $n$-band systems, making the plethora of exceptional structures deviate further from naive expectations.
We illustrate our findings in simple periodic toy models.
By relying on similarity relations instead of the less general symmetries, we simultaneously cover several physically relevant scenarios, ranging from optics and topolectrical circuits to open quantum systems.
This makes our predictions highly relevant and broadly applicable in modern research, as well as experimentally viable within various branches of physics.  
\end{abstract}
\maketitle

\section{Introduction}
\label{s1}
The past decade has marked the advent of non-Hermitian topological physics~\cite{Bergholtz2021}, expanding extensively on the previous utilization of non-Hermitian operators, which was largely focused on optics, where these operators are used to model gain and loss~\cite{ozdemir2019,Lu2014,Ozawa2019}. 
Although they violate fundamental laws of quantum mechanics, non-Hermitian operators may also serve as effective descriptions of open quantum systems~\cite{Khandelwal2021,Hatano2019,Lindblad1976}, ultracold atomic setups~\cite{Kreibich2014}, friction in mechanics~\cite{Ghatak2020}, resistivity in electrical circuits~\cite{Schindler2012,Yang2022}, and damping matrices in Liouvillian descriptions~\cite{Fan2022}, to mention a few examples.
The most profound differences from conventional Hermitian systems include the biorthogonal bulk-boundary correspondence~\cite{Kunst2018,Yao2018,Edvardsson2019}, the non-Hermitian skin effect~\cite{Yao2018, Kreibich2014, Gohsrich2025}, and the generic appearance of exceptional points (EPs) in the complex eigenvalue spectrum~\cite{Kato1966}.
The latter marks a simultaneous coalescence of eigenvalues and eigenvectors, leaving the corresponding operator at a Jordan normal form instead of a diagonal form~\cite{Kato1966}.
Remarkably, the generic appearance of $n$-fold EPs (EP$n$s) requires the tuning of $2n-2$ real parameters~\cite{Bergholtz2021,Heiss2012,Rotter2009,Berry2004, Delplace2021, Sayyad2022}, making them vastly more abundant than their Hermitian counterparts, which require the simultaneous tuning of $n^2-1$ [$2 (n^2-1)$] real parameters in (non-)Hermitian systems.
Consequently, EP2s appear in a stable fashion already in two dimensions, while three-dimensional (3D) systems may host embeddings of one-dimensional (1D) submanifolds of EP2s potentially taking the form of links~\cite{Carlstrom2018,Molina2018,Moors2019} and knots~\cite{Carlstrom2019,Zhang2021,Stalhammar2019,Wang2021}.
The number of tuning constraints is often referred to as the codimension of the corresponding degeneracy.

Focus has recently been directed toward further reducing the codimension of EP$n$s.
This is commonly done by exposing the dynamical matrix, the non-Hermitian operator governing the evolution of the system, to certain symmetries~\cite{Sayyad2022}; parity-time ($\mathcal{PT}$) symmetry is for instance known to reduce the codimension to $n-1$~\cite{Budich2019,Sayyad2022,Sayyad2023,Mandal2021,Delplace2021,Stalhammar2021}, while the codimension of EP$n$s induced by sublattice symmetry is $n-1$ ($n$) if $n$ is odd (even), thus making it depend on the parity of $n$~\cite{Delplace2021,Mandal2021,Sayyad2022}.
Mathematically speaking, all the different non-Hermitian symmetries that reduce the codimension of EP$n$s are merely special cases of one of three generalized similarity relations~\cite{Montag2024_2}.
Although the symmetries, their generators, and an appropriate basis must be chosen to physically interpret the EP$n$s, their topological properties  are captured already at the level of similarity relations, which are naturally carried over to the more physically relevant symmetries.

Even though stable EP$n$s (when $n>2$) always appear as special points on manifolds of EP$m$s, $m<n$, previous studies have almost exclusively treated EP$n$s as isolated objects, meaning that little attention has been directed toward the concomitant less degenerate exceptional structures.
Motivated by this, in this work we map out the different exceptional structures appearing in non-Hermitian systems subject to pseudo-(anti-)Hermiticity and self-skew similarity in three and four dimensions, respectively, which are the perfect examples to show the hierarchy of exceptional structures of various dimensions and various orders.
We find that the similarity-induced EP4s appearing in 3D pseudo-Hermitian four-band systems are not only accompanied by EP3 arcs and EP2 surfaces, but also by purely real or complex conjugate pairs of EP2 arcs.
The similarity-induced EP4s appearing in four-dimensional (4D) self-skew-similar four-band systems, on the other hand, are merely accompanied by two different EP2 surfaces, as the spectral constraint induced by the similarity forbids the emergence of EP3s of any kind.
This highlights the importance of not only reading off the number of constraints associated with a certain exceptional structure, but to further analyze how the similarity-induced spectral constraints can be satisfied.

When subjecting a non-Hermitian matrix to two different similarity relations simultaneously, the third one necessarily follows, and the codimension of EP$n$s is further reduced to $\lfloor \frac{n}{2} \rfloor$, with $\lfloor x \rfloor$ denoting the integer part of $x$~\cite{Montag2024}.
This induces even more exotic exceptional spectral features, which were recently uncovered for systems in two dimensions~\cite{Montag2024}.
In addition to the above, we here unravel the exceptional structures induced by multiple similarities in three and four dimensions.
We find EP6s and EP7s in 3D six- and seven-band models, respectively, while 4D eight- and nine-band models host EP8s and EP9s, respectively.
These pointlike structures are special points on a zoo of higher-dimensional manifolds of EPs of lower degeneracy, but, crucially, not all types of exceptional degeneracies are allowed by the similarity-induced spectral relations.
The Abelian eigenvalue topology of these pointlike EP$n$s is classified in terms of resultant winding numbers, complementing the recently developed classification scheme of multifold EPs~\cite{Yoshida2024,Stalhammar2024}.

Our work deepens the fundamental understanding of non-Hermitian topological band theory and multifold EP$n$s, and forms the final chapter in the study of analytically tractable similarity-induced exceptional structures \cite{Budich2019, Okugawa2019, Delplace2021, Mandal2021, Sayyad2022, Montag2024_2, Montag2024}. 
Our results can be used in several areas of modern physics research within both the classical and the quantum regime.
Significant examples include, but are not limited to, optics and topological photonics, where EPs can be directly mapped out using single-photon interferometry~\cite{Wang2021,Wang2023}, and ultracold atomic setups~\cite{Liu2024}.

The remainder of this article is organized as follows.
We set the stage in Sec.~\ref{s2} by showing how generalized similarities lower the codimension of EPs and induce intricate lower-order (less degenerate) exceptional structures between EP$n$s in $n$-band systems.
Further, we introduce generic minimal models and show that the nonvanishing terms in the characteristic polynomial and the spectral symmetry are equivalent to the generalized similarity constraints. 
Section~\ref{s3} provides the exceptional spectral structure connecting EP4s induced by pseudo-(anti-)Hermiticity in three dimensions, while Sec.~\ref{s4} does the same for self-skew-similar systems in four dimensions.
We address exceptional structures induced by multiple simultaneous similarities in three and four dimensions in Sec.~\ref{s5}.
We show the generic connection between EP$(2n)$s and EP$(2n+1)$s, and provide a classification of the Abelian eigenvalue topology of EP$n$s induced by multiple similarities in terms of resultant winding numbers.
We conclude and put our results in a wider perspective in Sec.~\ref{s6}.
The Supplemental Material~\cite{Supmat} devoted to technical results, where we prove in SM1 the equality between similarity relations and spectral symmetries, maps out the interesting Fermi structures accompanying the exceptional structures present in the models treated in the main text in SM2, and we fully uncover in SM3 the exceptional structures present in 4D systems subject to multiple similarities.

\section{Similarity-induced multifold exceptional points}
\label{s2}

Matrix similarity relations are generalizations of symmetry relations and are obtained by relaxing the unitarity constraint on symmetry generators.
Despite being more general, the similarities enforce spectral relations whose EP-inducing properties are the same as for the symmetries~\cite{kawabata2019,Sayyad2022,Montag2024_2}.
We devote this section to a brief survey on this topic, by introducing matrix similarities in Sec.~\ref{sec:2a} and their EP-inducing properties in Sec.~\ref{sec:2b}.
Section~\ref{sec:2c} shows that the exceptional structures induced by similarities can be illustrated using toy models derived from Frobenius companion matrices.

\subsection{Similarities in non-Hermitian systems} \label{sec:2a}

In previous work, it was established that unitary symmetries of non-Hermitian systems, which are local in parameter space, reduce the codimension of EP$n$s and that there are six such distinct symmetries~\cite{Sayyad2022}.
Recent work by some of the authors of this paper has shown that these six unitary symmetries are special cases of three similarities: pseudo-Hermiticity, pseudo-anti-Hermiticity, and self-skew-similarity~\cite{Montag2024_2}.
Each similarity is defined in terms of an invertible generator, as listed in Table~\ref{tab:sim_def}.
\begin{table}
    \centering
    \caption{Definitions of generalized self-similarities }.\label{tab:sim_def}
    \begin{tabular}{p{0.24\linewidth} p{0.38\linewidth} p{0.33\linewidth}}
        \hline \hline 
        Similarity  & Similarity constraint & Spectral constraint\\
        \hline
        \begin{tabular}{@{}l@{}}pseudo- \\ Hermiticity  \end{tabular} & $H(\bm{k}) = \eta H^\dagger(\bm{k}) \eta^{-1}$ & $\{\lambda(\bm{k})\} = \{\lambda^*(\bm{k})\}$ \\
        \begin{tabular}{@{}l@{}}pseudo anti- \\ Hermiticity  \end{tabular} & $H(\bm{k}) = -\Gamma H^\dagger(\bm{k}) \Gamma^{-1}$ & $\{\lambda(\bm{k})\} = \{-\lambda^*(\bm{k})\}$ \\
        \begin{tabular}{@{}l@{}}Self-skew \\ similarity  \end{tabular} & $H(\bm{k}) = -S H(\bm{k}) S^{-1}$ & $\{\lambda(\bm{k})\} = \{-\lambda(\bm{k})\}$ \\
        \hline \hline
    \end{tabular}
    {\raggedright Here, the operators $\eta,\Gamma$, and $S$ are invertible, and $\eta$ and $\Gamma$ are Hermitian. \par}
\end{table}
Different choices of generators amount to different allowed contributions to the dynamical matrix.
Their interpretation depends on the specific generator chosen, but we will not focus on this and rather work out general features induced by the similarities.
The generator is either constant, i.e., independent of the point in parameter space, in which case we call the similarity a global similarity, or the generator depends explicitly on the momentum $\bm{k}$ and the similarity is called parametric similarity.
All the following statements are true, regardless of whether a model has a global similarity or parametric similarity at every point in parameter space.

Each similarity is closely connected to a specific \emph{spectral symmetry} (cf. Table~\ref{tab:sim_def}), which follows immediately from the similarity relation.
By constructing a parametric similarity generator for a dynamical matrix with a specific spectral symmetry, we show in SM1~\cite{Supmat} that spectral symmetry implies similarity, where we make use of the results obtained in Refs.~\cite{Zhang2020b,Montag2024_2}.
This results in the following statement: Iff a non-Hermitian finite-dimensional dynamical matrix fulfills the pseudo-Hermitian, anti-pseudo-Hermitian, or self-skew-similar similarity constraint, it has the spectral symmetry $\{\lambda(\bm{k})\} = \{\lambda^*(\bm{k})\}$, $\{\lambda(\bm{k})\} = \{-\lambda^*(\bm{k})\}$, or $\{\lambda(\bm{k})\} = \{-\lambda(\bm{k})\}$, respectively, and vice versa.
These spectral symmetries play a key role in the appearance of multifold EPs in lower dimensions.

\subsection{Similarity-induced exceptional points} \label{sec:2b}

To see how the similarities induce stable multifold EPs, we introduce the conditions for them to appear in generic systems.
An $n$-band system exhibits an EP$n$ if and only if all terms except the leading one in the characteristic polynomial vanish.
The coefficients of the characteristic polynomial can be expressed as sums of the determinant $\det[H]$ and $n-2$ different traces $\tr[H^k]$, with $k=2,\dots,n-1$~\cite{Curtright2012}. 
We set $\tr[H]=0$ without loss of generality throughout, since a finite trace of a dynamical matrix $H$ does not change its rank.
Both the determinant and the $n-2$ traces are generally complex for non-Hermitian systems.
This amounts to $2(n-1)$ real constraints, which need to be simultaneously enforced to find an EP$n$~\cite{Sayyad2022}.
The number of constraints is referred to as the codimension of the EP$n$.

Due to the spectral symmetries imposed by similarities, some of the constraints defining the emergence of EP$n$s trivially hold as a consequence of the similarity.
Thus, the codimension of EP$n$s can be reduced by imposing generalized self-similarities on the system~\cite{Montag2024_2}.
The remaining constraints for $n$-band systems are derived in Refs.~\citenum{Sayyad2022} and \citenum{Montag2024_2}, and are listed in Table~\ref{tab:similarities}.

We emphasize that any unitary non-Hermitian symmetry enforcing spectral symmetry is a special case of one of the similarities~\cite{Montag2024_2}.
Any statement holding for similarity constrained non-Hermitian systems holds for a non-Hermitian system constrained by such a unitary symmetry.
The multifold exceptional structures emerging in two dimensions have been systematically studied in Ref.~\cite{Montag2024} and amount to EP3s, EP4s, and EP5s induced by unitary symmetries. 
The results generalize directly to the similarity-induced EP$n$s in two dimensions, because the symmetries constitute special cases of those similarities.

\begin{table*}
    \centering
    \caption{Number of constraints for realizing EP$n$s in $n$-band systems restricted by generalized self-similarities and constraints on the terms in the characteristic polynomials for similarity constrained models.}\label{tab:similarities}
    \begin{tabular}{p{0.23\linewidth}|p{0.18\linewidth}|p{0.18\linewidth}|p{0.18\linewidth}|p{0.18\linewidth}}
        \hline \hline 
        Similarity & \multicolumn{2}{l|}{$\#$ Constraints/codimension} & \multicolumn{2}{l}{Coefficients of characteristic polynomial} \\ \cline{2-5}
        $[\textrm{spectrum}]$ & $n \in $ even & $n \in$ odd & $j=0 \vee j \in $ even & $j \in$ odd \\
        \hline \hline
        \begin{tabular}{@{}l@{}}Pseudo-Hermiticity \\ $[\{\lambda\}=\{\lambda^*\}]$ \end{tabular} & $n-1$~
        $\begin{cases}
            \Re[\det({\cal H})] , \\
            \Re[\tr({\cal H}^k)] 
        \end{cases}$  & $n-1$~
       $\begin{cases}
            \Re[\det({\cal H})], \\
            \Re[\tr({\cal H}^k)] 
        \end{cases}$ & $a_j \in \mathbb{R}$  & $a_j \in \mathbb{R}$ \\
        \hline
        \begin{tabular}{@{}l@{}}Pseudo-anti-Hermiticity \\ $[\{\lambda\}=\{-\lambda^*\}]$ \end{tabular} & $n-1
        \begin{cases}
            \Re[\det({\cal H})] , \\
            \Re[\tr({\cal H}^{l})] ,\\
            \Im[\tr({\cal H}^{m})] 
        \end{cases}$  & $n-1$~
       $\begin{cases}
            \Im[\det({\cal H})] , \\
            \Re[\tr({\cal H}^{l})] , \\
            \Im[\tr({\cal H}^{m})] 
        \end{cases}$ &
        $a_j
        \begin{cases}
            \in \mathbb{R} \; \text{if} \; n \in \text{even} , \\
            \in i \mathbb{R} \; \text{if} \; n \in \text{odd}
        \end{cases}$  & $a_j
        \begin{cases}
            \in i \mathbb{R} \; \text{if} \; n \in \text{even} , \\
            \in \mathbb{R} \; \text{if} \; n \in \text{odd}
        \end{cases}$ \\
        \hline 
        \begin{tabular}{@{}l@{}}Self-skew-similarity\\ $[\{\lambda\}=\{-\lambda\}]$ \end{tabular} & $n
        \begin{cases}
            \det({\cal H}) , \\
            \tr({\cal H}^{l}) 
        \end{cases}$  & $n-1$~
       $\begin{cases}
            \tr({\cal H}^{l}) 
        \end{cases}$ & $a_j
        \begin{cases}
            \in \mathbb{C} \; \text{if} \; n \in \text{even} , \\
            =0 \; \text{if} \; n \in \text{odd}
        \end{cases}$  & $a_j
        \begin{cases}
            =0 \; \text{if} \; n \in \text{even} , \\
            \in \mathbb{C} \; \text{if} \; n \in \text{odd}
        \end{cases}$ \\
        \hline 
        \begin{tabular}{@{}l@{}}Combined$^1$ \\ $[\{\lambda\}=\{\lambda^*\} \, \wedge \, \{\lambda\}=\{-\lambda\}]$ \end{tabular} & $\frac{n}{2}
        \begin{cases}
            \Re[\det({\cal H})] ,\\
            \Re[\tr({\cal H}^{l})] 
        \end{cases}$  & $\frac{n-1}{2} 
        \begin{cases}
            \Re[\tr({\cal H}^{l})] 
        \end{cases}$ & $a_j
        \begin{cases}
            \in \mathbb{R} \; \text{if} \; n \in \text{even} , \\
            =0 \; \text{if} \; n \in \text{odd}
        \end{cases}$  & $a_j
        \begin{cases}
            =0 \; \text{if} \; n \in \text{even} , \\
            \in \mathbb{R} \; \text{if} \; n \in \text{odd}
        \end{cases}$ \\
        \hline \hline
    \end{tabular}
    {\raggedright Here, $k\in \{2, \dots n-1\}$, $l \in \{2 \leq l < n, l\in\textrm{even} \}$ and $m \in \{3 \leq m < n, m\in\textrm{odd} \}$. Behind the number of constraints, we write the specific quantities that need to be set to zero to find EP$n$s. $^1$Here, combined encompasses the constraints enforced by any pair of similarities above.\par}
\end{table*}

\subsection{Minimal Frobenius companion models} \label{sec:2c}

The exceptional structures induced by similarities are highlighted by toy models below.
Minimal models with a certain similarity will be constructed using the Frobenius companion matrix
\begin{equation}
    H = \begin{pNiceMatrix}
             0 & 1 &0 & \Cdots & 0   \\
             0      & \Ddots &  \Ddots & \Ddots & \Vdots \\
             \Vdots & \Ddots &&1& 0  \\
             0& \Cdots &0&0& 1\\
             a_0      & a_1 & \Cdots & a_{n-2} & 0  \\
          \end{pNiceMatrix} \, ,
\end{equation}
where $a_i\in\mathbb{C}$ in general~\cite{Brand1964,Arnold1971,Yoshida2024,Sayyad2022}.
The characteristic polynomial is given by 
\begin{equation}
    P_n(\lambda) = (-1)^n \, \left(\lambda^n - \sum_{j=0}^{n-2} a_j \lambda^j\right) \, .
\end{equation}
The spectral symmetries enforced by the similarities impose constraints on $a_i$, which are listed in Table~\ref{tab:similarities}.
Thus, by imposing the constraints from Table~\ref{tab:similarities} on the models we obtain toy models for all possible similarities according to the proof in SM1 in the Supplemental Material~\cite{Supmat}.

\section{Exceptional Structures Induced by Pseudo-Hermiticity}
\label{s3}
EP$n$s induced by pseudo-Hermiticity have hitherto to a large extent been studied as isolated objects, thus neglecting (or avoiding) the fact that these appear as special points of manifolds of less degenerate EPs.
In this section, we map out these exceptional structures, initially in 3D (Sec.~\ref{sec:EP3D}), and include an example in Sec.~\ref{sec:phex} before general conclusions are made (Sec.~\ref{sec:EPGenPsH}). We comment on pseudo-anti-Hermiticity in Sec.~\ref{sec:pah}.

\subsection{General considerations} \label{sec:EP3D}

It is known that EP4s emerge pairwise and in a stable fashion in pseudo-Hermitian four-band systems in a 3D parameter space, and a topological classification of them was recently derived~\cite{Yoshida2024}.
Here, we study the exceptional structures between them and show that they comprise EP2 surfaces, special arcs on those surfaces, and EP3 arcs connecting pairs of EP4s.

Let us show this in detail. 
Consider a general pseudo-Hermitian four-band dynamical matrix $H(\bm{k}) = \eta H^\dagger (\bm{k}) \eta^{-1}$, shift it by a $\bm{k}$-dependent constant such that $\tr[H(\bm{k})]=0$, and introduce
\begin{align}
    \alpha &= -\frac{1}{6} \tr\left(H^2\right) \, , \\
    \beta &= -\frac{1}{3\sqrt{2}} \tr\left(H^3\right) \, ,
    \\
    \gamma &= \frac{4}{3} \det\left(H\right) \, .
\end{align}
Using the pseudo-Hermiticity constraint, we find that all three parameters $\alpha,\beta,$ and $\gamma\in\mathbb{R}$~\cite{Sayyad2022, Montag2024_2} and thus the characteristic polynomial is a real fourth-order polynomial given by
\begin{equation}
    \mathcal{P}_4 (\lambda) = \lambda^4 + 3 \alpha \lambda^2 + \sqrt{2} \beta \lambda + \frac{3}{4} \gamma \, .
\end{equation}
The third-order term vanishes, because we are only considering traceless dynamical matrices~\cite{Sayyad2022}.
The discriminant of this polynomial is given by
\begin{equation}
    \begin{split}
        \mathcal{D}_4 = &-108 \, \left( 2\alpha^3\beta^2 + \beta^4 - 9 \alpha^4 \gamma\right. \\
        &- \left. 6 \alpha \beta^2 \gamma + 6 \alpha^2\gamma^2 - \gamma^3 \right) \, .
    \end{split}
\end{equation}
The eigenvalues of a real fourth-order characteristic polynomial are either real or appear as complex conjugate pairs, thus fulfilling the spectral symmetry.

Since EPs appear at eigenvalue degeneracies, they correspond to points where the discriminant vanishes.
Strictly speaking, this is only a condition for finding spectral degeneracies, but since diabolic points, i.e., nondefective degeneracies or ``Hermitian'' degeneracies, require more constraints to be fulfilled [$2(n^2 - 1)$ real constraints] they have higher codimension~\cite{Sayyad2023}.
It is known that similarity-induced EP4s appear if and only if $\alpha=\beta=\gamma=0$~\cite{Sayyad2022,Montag2024_2}.
When these constraints are met, they individually define closed surfaces in the periodic 3D parameter space, whose intersection are generically points appearing pairwise.
This can be visualized by first considering the intersection between two of the three closed surfaces, which is a closed loop.
If this loop intersects with the third surface, there must be at least two points where all three constraints are fulfilled.
Any additional intersection of the loop with the surface will lead to the emergence of another pair of EP4s.
Apart from this geometric argument, a doubling theorem based on the resultant winding number of the EP4, generalizing to general EP$n$s in $n$-band systems, has recently been derived~\cite{Yoshida2024}.

Similar to similarity-induced EP3s in 2D parameter spaces, we expect the EP4 pairs to be connected by exceptional structures of lower degeneracy.
By rewriting the discriminant and setting it to zero, these structures arise as solutions to
\begin{equation} \label{eq:DiscPSH}
    \beta^2\left[\beta^2+2\left(\alpha^3-3\alpha\gamma\right)\right] - \gamma \left(3\alpha^2-\gamma\right)^2 = 0 \, ,
\end{equation}
which results in the following:
\begin{figure*}[htb!]
\includegraphics[width=\textwidth]{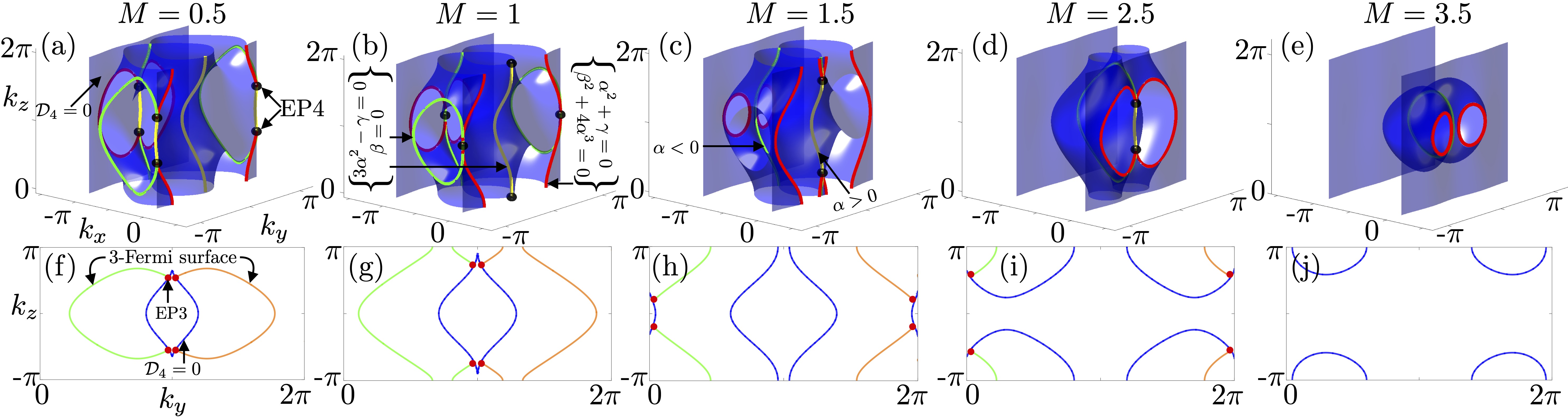}
\caption{Exceptional structures of the pseudo-Hermitian four-band model in 3D given by Eqs.~\eqref{eq:PSHHam1}-\eqref{eq:PSHHam3} in panels (a)-(e), and their concomitant three-level Fermi surfaces in for $k_x=0$ in panels (f)-(j). In panels (a)-(e), the blue surfaces are similarity-induced EP2s ($\mathcal{D}_4=0$), the red arcs are similarity-induced EP3s ($\alpha^2+\gamma=\beta^2+4\alpha^3=0$), and the black dots are similarity-induced EP4s $(\alpha=\beta=\gamma=0$). The green and yellow arcs, however, denote special EP2s of codimension 2. The yellow (green) arcs arise as solutions to $\beta=3\alpha^2-\gamma=0$ with $\alpha>0$ ($\alpha<0$). Panels (a)-(e) display how the EP4s are pairwise created/annihilated as a function of the real parameter $M$, a process that affects the appearance of the similarity-induced EP3 arcs, but also the special EP2 arcs. In panels (f)-(j), the behavior of the three-level Fermi surfaces (green and orange arcs), where three of the four eigenvalues share identical real parts, and the discriminant (blue arc) is displayed. The green and orange arcs correspond to three-level Fermi surfaces given by Eqs.~\eqref{eq:3FS1} and \eqref{eq:3FS2}, respectively. For illustrative purposes, this is displayed for a 2D cut of the Brillouin zone given by $k_x=0$. The intersection of the three-level Fermi surfaces with the discriminant forms EP3s (red dots), and when there are no intersections, the EP3s are gapped out, as shown in panel (j).
\label{fig:EPPSH} }
\end{figure*}

\begin{itemize}
    \item[(1)] Surfaces of real EP2s given by $\mathcal{D}_4=0$ for $\beta\neq0$, accompanied by two nondegenerate real or a pair of complex conjugate eigenvalues;
    \item[(2)] arcs along which pairs of purely real EP2s appear at the intersection of different EP2 surfaces. Both EP2s are similarity induced, but them appearing simultaneously results in the codimension of 2 associated with this structure;
    \item[(3)] arcs of real EP3s emerging on the EP2 surfaces;
    \item[(4)] arcs of complex conjugated pairs of purely imaginary EP2s appear connected to the remaining exceptional structure only at the EP4s. Their emergence is again not similarity induced, and the similarity constraint instead results in the pairwise appearance.
\end{itemize}
Let us explain in detail how these EPs emerge in the following.

The real EP2s (1) emerge on two closed surfaces, where the isolated surfaces are defined by Eq.~(\ref{eq:DiscPSH}) for $\beta\neq0$.
The intersection of these surfaces, corresponding to $\beta=0$ and $3\alpha^2-\gamma=0$ for $\alpha<0$, defines a closed loop on which there is a pair of real EP2s (2).
Here, two constraints need to be fulfilled, which can be interpreted as one constraint per EP2. 
This loop connects to the EP4s at $\alpha=\beta=\gamma=0$.
The dimension of the real EP2 surface agrees with the expected dimensions from the codimension of EP2s in pseudo-Hermitian systems.

EP3s (3) appear if $\alpha^2+\gamma=0$ and $\beta^2+4\alpha^3=0$.
These EP3 arcs emerge on the real EP2 surfaces.
EP4 pairs are connected by two topologically distinct EP3 arcs.
These are distinguished by the sign of the degenerate eigenvalue, or, equivalently, the sign of $\beta$.
The codimension of the EP3 arcs is reflected in the number of constraints that must be fulfilled for the emergence of the EP3 arcs in pseudo-Hermitian systems.

Standing out slightly are the pairs of imaginary EP2 arcs (4). 
These emerge for $\beta=0$ and $3\alpha^2-\gamma=0$, with $\alpha>0$, also connecting the EP4s at $\alpha=\beta=\gamma=0$.
The EP2s on these arcs have codimension 2, because two constraints need to be fulfilled for them to appear.
Therefore, they have a higher codimension than we expect of EP2s in pseudo-Hermitian systems.
The reason for this is that these EP2 arcs are not similarity induced, but rather ordinary EP2s with codimension 2.
Instead of reducing the codimension, the spectral symmetry forces these imaginary EP2 arcs to come in complex conjugate pairs, leaving them to necessarily emerge with codimension 2.

No other EPs can emerge in the presence of pseudo-Hermiticity in a four-band system.
In addition to the exceptional structures, pseudo-Hermitian systems exhibit interesting Fermi structures, which in the case of four-band models form three-level Fermi surfaces.
At these, three of the four eigenvalues share the same real part while still being nondegenerate, a condition for which a closed-form expression can be derived analytically.
By introducing $u,v$, and $w$ as,
\begin{align}
\alpha &= \frac{1}{3}\left(v+w-4u^2\right) \, ,
\\
\beta &= \sqrt{2} u \left(w-v\right) \, ,
\\
\gamma&=\frac{4}{3}wv \, ,
\end{align}
the characteristic polynomial can, by using Descartes method, be recast into the form
\begin{equation}
    \mathcal{P}_4(\lambda) = \left(\lambda^2 + 2u\lambda +v \right)\left(\lambda^2-2u\lambda+w\right)\,.
\end{equation}
The eigenvalue solutions $\mathcal{P}_4(\lambda_{i, \pm}) = 0$ then read
\begin{equation}
    \lambda_{1,\pm}=u\pm \sqrt{u^2-v}, \quad \lambda_{2,\pm} = -u\pm \sqrt{u^2-w},
\end{equation}
giving three-level Fermi surfaces when the following conditions are met:
\begin{align}
    4u^2&=u^2-v, \quad u^2-w<0 \, , \label{eq:3FS1}
    \\
    4u^2&=u^2-w, \quad u^2-v<0 \,. \label{eq:3FS2}
\end{align}
These are of importance since their intersections with the EP2 surface exactly form the EP3 arcs.
The full Fermi structure of pseudo-Hermitian four-band systems is highlighted in SM2 in the Supplemental Material~\cite{Supmat}.

\subsection{An example} \label{sec:phex}

To illustrate the exceptional spectral features, we consider the following example of a pseudo-Hermitian four-band model in a 3D parameter space given by
\begin{align}
    H_4 &= \begin{pmatrix}0&1&0&0\\0&0&1&0\\0&0&0&1\\a_0 & a_1& a_2&0 \end{pmatrix}\, \label{eq:PSHHam1},
    \\
    a_0&=\sin(k_x),\quad a_1=\sin(k_y), \label{eq:PSHHam2}
    \\
    a_2 &=M-\cos(k_x)-\cos(k_y)+\cos(k_z). \label{eq:PSHHam3}
\end{align}
$k_x,k_y,$ and $k_z$ denote the lattice momentum components, and $M$ is a real constant parameter. 
This model is inspired by the (Hermitian) Weyl semimetal model used in Ref.~\cite{Udagawa2016}, and its eigenvalue structures are shown in Fig.~\ref{fig:EPPSH}. 
Figures~\ref{fig:EPPSH}(a)-(e) show how the surfaces of similarity-induced EP2s (blue surfaces), the similarity-induced arcs of EP3s (red arcs), and the special EP2 arcs (yellow arcs are complex conjugated, and green arcs, corresponding to the intersection between the EP2 surfaces, are fully real) are all connected to the similarity-induced EP4s (black dots) in the way predicted above.
Moreover, Figs.~\ref{fig:EPPSH}(f)-(i) show the three-level Fermi surfaces [green arc for Eq.~\eqref{eq:3FS1} and orange arc for Eq.~\eqref{eq:3FS2}] and the vanishing of the discriminant (blue arc) in the cut $k_x=0$.
As claimed, their intersection forms EP3s (red dots), and when their intersection is empty, no EP3s appear in the spectrum.

\subsection{Dimensional generalization} \label{sec:EPGenPsH}
From the observations made in the special case presented in the previous subsection, we see that the pseudo-Hermitian similarity-induced EP structures are accompanied by some generic exceptional structures. 
Using these insights, we here identify the generic EP structures in the general case. 
To do this, we will treat the cases of even and odd bands separately.

When the matrix dimension $n$ is even, a pseudo-Hermitian spectrum is achieved by an operator of the form
\begin{equation}
    H = \begin{pmatrix} \mathcal{H} &0\\0&\mathcal{H}^{\dagger} \end{pmatrix}
\end{equation}
for some generic dynamical $n/2 \times n/2$ matrix $\mathcal{H}$ and its conjugate transpose $\mathcal{H}^{\dagger}$.
Since the corresponding spectrum satisfies the symmetries induced by pseudo-Hermitian similarity, we can by the proof in SM1 in the Supplemental Material~\cite{Supmat} conclude that such a matrix is pseudo-Hermitian with respect to some generator.
Since $\mathcal{H}$ is a generic matrix, i.e., it does not fulfill any similarity constraint, the exceptional structures appearing blockwise will not have a reduced codimension, but rather appear as generic exceptional structures.
The spectral constraint induced by the similarity will instead ensure that these structures emerge either as complex conjugate paired structures or as purely real structures; they either appear at purely real eigenvalues or they appear pairwise in each half of the complex eigenvalue plane.
This means that in addition to the similarity-induced structures of EP$k$s of codimension $k-1<n$, there will be generic structures of EP$m$s with $m\leq n/2$ of the generic codimension $2m-2$, which either are restricted to the real eigenvalue axis or emerge at complex conjugate eigenvalues.

When $n$ is odd, a pseudo-Hermitian spectrum is instead achieved by an operator on the form
\begin{equation}
    H = \begin{pmatrix} \mathcal{H} &0&0\\0&\mathcal{H}^{\dagger} &0\\0&0&\lambda\end{pmatrix},
\end{equation}
where $\mathcal{H}$ is again an $(n-1)/2\times (n-1)/2$ matrix representation of a generic operator $\mathcal{H}$, and $\mathcal{H}^{\dagger}$ its conjugate transpose. 
Here, $\lambda$ is necessarily real, as it comprises a pseudo-Hermitian eigenvalue whose conjugate does not appear in the rest of the spectrum.
Here, the same reasoning as for the case of an even number of bands holds, meaning that the similarity-induced structures of EP$k$s of codimension $k-1<n$ will be accompanied by generic structures of EP$m$s with $m\leq(n-1)/2$ of codimension $2m-2$, emerging at purely real or at complex conjugate eigenvalues.

\begin{figure}
\includegraphics[width=0.85\columnwidth]{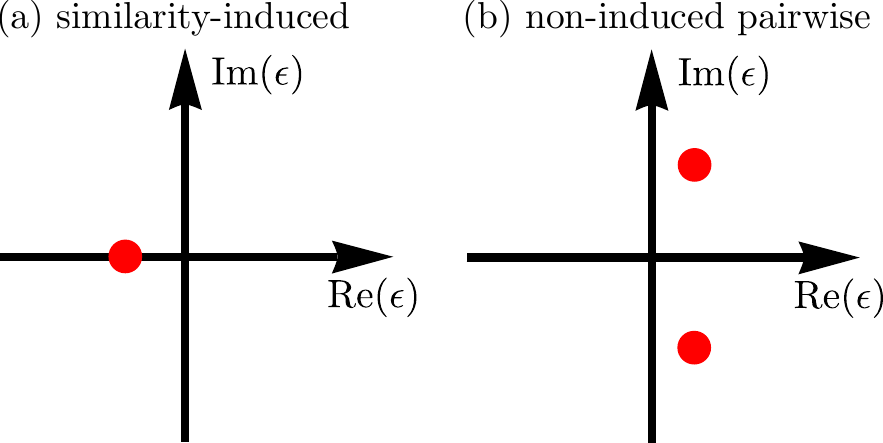}
\caption{Illustration of how pseudo-Hermitian similarity constrains the position of exceptional structures of $n$-band systems (red dots) in the complex eigenvalue plane. (a)~Similarity-induced EP$m$s of codimension $m-1$ are, when $m \leq n$, forced to appear at real eigenvalues. (b)~Despite the similarity, EP$m$s of order $m \leq \lfloor \frac{n}{2} \rfloor$, with $\lfloor x \rfloor$ denoting the integer part of $x$, of generic codimension $2m-2$ may still appear in the spectrum, but they are constrained to appear pairwise at complex conjugate eigenvalues. 
\label{fig:Loc_PsH} }
\end{figure}

To conclude, the spectrum of a pseudo-Hermitian operator not only hosts similarity-induced EP$k$s of codimension $k-1$, but also generic EP$m$s of codimension $2m-2$ when the number of bands is at least $2m$.
Figure~\ref{fig:Loc_PsH} illustrates that these EP$l$s are constrained to emerge at the real axis and at complex conjugate eigenvalues, respectively,  to assure them fulfilling the similarity-induced spectral constraints.

\subsection{Pseudo-anti-Hermiticity} \label{sec:pah}

The results in this section can be extended to pseudo-anti-Hermitian systems in a straightforward manner:
The exceptional structures for a pseudo-anti-Hermitian system $\Tilde{H}(\bm{k})$ are found by rotating those of a pseudo-Hermitian system $H'=-i\Tilde{H}$ by $\pi/2$ in the complex plane. This means that any eigenvalue $\lambda$ of $H'$ is mapped to $\Tilde{\lambda}=-i\lambda$, which is an eigenvalue of $\Tilde{H}$.
This shows that any exceptional point in the spectrum of $\Tilde{H}$ can be derived by considering the exceptional structure of $H'$.
For four-band pseudo-anti-Hermitian models in particular, this implies the presence of pairs of EP4s, connected by surfaces of imaginary EP2s, intersecting in lines of paired imaginary EP2s and with EP3 lines emerging on them, and additional isolated arcs of noninduced paired real EP2s.

\section{self-skew-similarity-induced EPns} \label{s4}

Self-skew-similarity is different from the other two similarities, because it relates a dynamical matrix to itself and not its adjoint.
This results in a unique structure of the similarity-constrained characteristic polynomial.
Since it is the generalization of the (non-)Hermitian sublattice symmetry, which is present in many tight-binding models, self-skew similarity is of special importance.
In these systems, EP$n$s emerge pairwise in $n$ ($n-1$) dimensions when $n$ is even (odd), and they are topologically classified in Ref.~\cite{Stalhammar2024}.
Here, we show that the corresponding eigenvalues can be derived by mapping $2n$- and $(2n+1)$-band self-skew-similar models onto unconstrained $n$-band non-Hermitian models (Secs.~\ref{sec:2n-2n+1} and \ref{sec:mapping}).
We highlight with an example how the exceptional structure of self-skew-similar systems can be obtained with this method (Sec.~\ref{sec:sssmodel}).

\subsection{Likeness of \texorpdfstring{$\bm{2n}$}{2n} and \texorpdfstring{$\bm{2n+1}$}{2n+1} models} \label{sec:2n-2n+1}

\begin{figure*}[htb!]
\includegraphics[width=\textwidth]{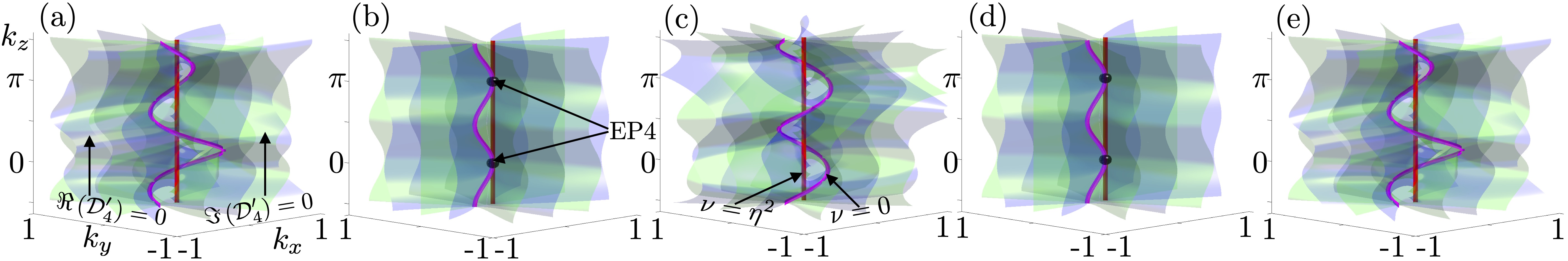}
\caption{Exceptional structures in part of the Brillouin zone of the model given by Eqs.~\eqref{eq:SSSHam1}-\eqref{eq:SSSHam3}. The blue (green) surfaces correspond to the real (imaginary) part of the discriminant $\mathcal{D}_4'$ of $P_4'(\lambda)$ being zero, while the red (magenta) arcs illustrate the EP2 2D surfaces in 4D given by $\nu=\eta^2$ ($\nu=0$) as 1D arcs in 3D. Panels (a)-(e) show different cuts in $k_w$, given by $k_w = \left(-\frac{3k_0}{2},-k_0,0,k_0,\frac{3k_0}{2}\right)$, respectively, with $k_0 = \cos^{-1}\left[\sum_{i=x,y,z}\cos(k_i)-M\right]$. When $k_w=-k_0$ [panel (b)] and $k_w=k_0$ [panel (d)], the system hosts EP4s appearing as intersections between the different EP2 structures, which do not appear when $k_w=-\frac{3k_0}{2}$ [panel (a)], $k_w = 0$ [panel (c)], or $k_w = \frac{3k_0}{2}$ [panel (e)]. 
Notable is further that the direction of winding of the magenta arc around the red arc changes when passing through the EP4s, something that can only occur through such a transition. 
\label{fig:EPSSS} }
\end{figure*}

The characteristic polynomials of a self-skew-similar system with $2n$ and $2n+1$ bands are given by
\begin{align}
    &P_{2n}(\lambda) = \left(\lambda^{2n} - \sum_{j=0}^{n-1} a_{2j} \lambda^{2j}\right) \, , \\
    &P_{2n+1}(\lambda) = -\lambda \left(\lambda^{2n} - \sum_{j=0}^{n-1} a_{2j+1} \lambda^{2j}\right) \, .
\end{align}
For an odd number of bands, $\lambda=0$ is thus always an eigenvalue of the system, appearing as a flat band.
This implies the existence of compact localized nondecaying states.
The form of the characteristic polynomial further makes it possible to achieve the spectral structure of a self-skew-similar $(2n+1)$-band system by adding a flat band to the spectrum of a self-skew-similar $2n$-band system by letting $a_{2j+1}\rightarrow a_{2j}$.
We find similarity-induced EP$2n$s at $\lambda=0$ in $2n$ dimensions, and since the flat band in the $(2n+1)$-band system is also at $\lambda=0$ the EPs are promoted to EP$(2n+1)$s.
The same argument holds for any exceptional structure of lower degeneracy at $\lambda=0$:
Any $k$-fold ($k\leq 2n$) exceptional structure at $\lambda=0$ found in $2n$-band systems in the presence of self-skew-similarity is promoted to a $(k+1)$-fold exceptional structure in $(2n+1)$-band systems.
Further, all exceptional structures found in $2n$-band systems at $\lambda\neq 0$ remain unaltered upon adding the $(2n+1)$th band.
Therefore, studying the exceptional structures of self-skew-similar $2n$-band systems also reveals the exceptional structure of self-skew-similar systems with $2n+1$ bands, which is why we from here on will consider self-skew-similar systems with an even number of bands exclusively.

\subsection{Mapping to unconstrained lower-band model} \label{sec:mapping}
The form of the characteristic polynomial of $2n$–band self-skew-similar systems can be mapped to that of a generic $n$-band system. 
Such a mapping will facilitate solving for the eigenvalues, and thus also finding the exceptional structures connected to the EP$2n$s. 
With the substitution
\begin{equation}
    z = \lambda^2 - \frac{a_{2n-2}}{n}
\end{equation}
the characteristic polynomial can be transformed to
\begin{equation}\label{eq:red_pol_SSS}
    \Tilde{P}_n(z) = z^n - \sum_{j=0}^{n-2} b_j z^j \, ,
\end{equation}
where the coefficients $b_j$ are complex-valued functions of $a_0,\dots,a_{2n-2}$.
Given the solution for the $n$ eigenvalues $z_l$, and by utilizing 
\begin{equation}
\lambda_\pm(z_l) = \pm \sqrt{z_l +a_{2n-2}/n},
\end{equation}
the eigenvalues of the initial self-skew-similar $2n$-band model can be derived.
An identical principle can be used for systems subject to multiple similarities, which we will see in Sec.~\ref{s5}.

\subsection{An example} \label{sec:sssmodel}
To exemplify the mapping and to highlight how the exceptional structure of higher-band self-skew-similar systems can be solved using this method, we consider a self-skew-similar four-band model described by 
\begin{align}
H' &= \begin{pmatrix} 0&1&0&0\\0&0&1&0\\0&0&0&1\\a_0&0&a_2&0\end{pmatrix}, \label{eq:SSSHam1}
\\
a_0&=\sin(k_x)+i\sin(k_y), \label{eq:SSSHam2}
\\
a_2 &= \sin(k_z)+i\left[M-\sum_{i=x,y,z}\cos(k_i)+\cos(k_w)\right], \label{eq:SSSHam3}
\end{align}
in a 4D space parametrized by $(k_x,k_y,k_z,k_w)$.
As shown before, all results obtained here can be transferred to a self-skew-similar five-band model by adding a flat band at the origin.
Introducing
\begin{align}
    \eta = -\frac{\tr[(H')^2]}{4}, \quad  \nu = \det[H'] -\left\{\frac{\tr[(H')^2]}{4}\right\}^2 \, ,
\end{align}
we can write the characteristic polynomial as
\begin{equation}
    \mathcal{P}'_4(\lambda) = \lambda^4 + 2\eta\lambda^2 + \nu+\eta^2 \, .
\end{equation}
By substituting $z=\lambda^2+\eta$, the reduced polynomial is given by
\begin{equation}
   \Tilde{\mathcal{P}}(z) = z^2 + \nu \, .
\end{equation}
Similarity-induced EP4s are then found when $\eta=\nu=0$.
These EP4s are connected by different surfaces of EP2s, given by (1) $\nu=0$, $\eta\neq 0$, and (2) $\nu=\eta^2$.
The constraint $\nu=0$ gives a pair of EP2 surfaces, connected at the EP4, while $\nu=\eta^2$ gives an EP2 surface appearing at $\lambda=0$, accompanied by two non-degenerate eigenvalues.
Figure~\ref{fig:EPSSS} illustrates how these exceptional structures appear in different 3D slices.
Consequently, the EP2s appear as arcs rather than surfaces, and the EP4s only appear in panels (b) and (d), corresponding to the slices in $k_w$ including the EP4s.
The fact that the EP2 arcs do not vanish regardless of the chosen $k_w$ slice, while the EP4s do, indicate that the EP2 structure is indeed a 2D surface embedded in 4D, while the EP4s are points in 4D. 

These are the only exceptional structures emerging between the induced EP4s.
Even though EP3s are expected to emerge in four dimensions, the self-skew-similarity prevents this for four-band systems; with four bands, it is impossible to have an EP3 and fulfill the spectral symmetry enforced by self-skew-similarity.
In addition to the exceptional structure, we find Fermi structures for these systems, which are described in SM2 in the Supplemental Material~\cite{Supmat}.

Finally, let us comment on how these exceptional structures are affected when adding a fifth band.
Following the similarity constraints, this band is necessarily flat and located at $\lambda = 0$.
Thus, the EP4s are promoted to EP5s.
Since the surfaces of EP2s appear at both finite and zero $\lambda$, they are affected differently upon adding the flat band.
The EP2 surface at $\nu=0$ becomes a 2D surface of EP2s embedded in 4D, while the EP2 surface at $\nu=\eta^2$ becomes a 2D surface of EP3s. 

\subsection{Dimensional generalization}
\begin{figure}
\includegraphics[width=0.85\columnwidth]{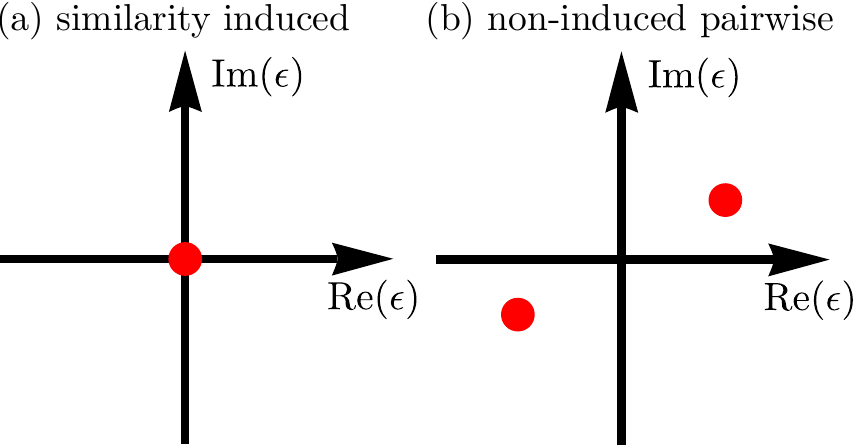}
\caption{Illustration of how self-skew-similarity constrains the position of exceptional structures of $n$-band systems (red dots) in the complex eigenvalue plane. (a) Similarity-induced EP$m$s are, when $m<n$, forced to appear at the origin and have codimension $2k$ when $m=2k+\delta$, where $\delta=0$ ($\delta=1$) for $n\in\text{even}$ ($n\in\text{odd}$). (b) Despite the similarity, EP$m$s of generic codimension $2m-2$ may still appear in the spectrum when $m<\frac{n}{2}$, but they are constrained to appear pairwise in adjacent quadrants of the complex eigenvalue plane.
\label{fig:Loc_SSS}}
\end{figure}
Just as for the pseudo-Hermitian systems treated in Sec.~\ref{s3}, the exceptional structures induced by self-similarity will be accompanied by generic exceptional structures.
The emergence of the generic exceptional structures can be argued for by studying a model given by
\begin{equation}
    H = \begin{pmatrix} \mathcal{H} &0\\0& -\mathcal{H}\end{pmatrix}.
\end{equation}
Since this model has a self-skew-similar spectrum, it is self-skew-similar with respect to some generator.
Here, $H$ is taken to be $n\times n$ with $n\in\text{even}$.
Since $\mathcal{H}$ is a completely generic non-Hermitian operator, the corresponding exceptional structures will be generic, which means that EP$m$s with $m\leq n/2$ are of codimension $2m-2$.
This means that the similarity-induced EP$k$s of codimension $k$ (recalling that $H$ models a system with an even number of bands) are accompanied by generic EP$m$s of codimension $2m-2$.
Figure~\ref{fig:Loc_SSS} illustrates that these are further constrained to appear at zero eigenvalue, or pairwise in adjacent quadrants of the complex eigenvalue plane, respectively, to ensure that the similarity-induced spectral constraints hold.

Upon adding a band to $H$, which by the similarity is forced to be flat appearing at $\lambda=0$, the similarity-induced EP$k$s will be of codimension $k-1$, and all exceptional degeneracies appearing at $\lambda=0$ will be increased by one.
Apart from that, the same reasoning as for an even number of bands holds. 

This concludes the analysis of self-skew-similar systems from the point of view of multifold exceptional structures. 
In the following section, we apply similar methods to study EP$n$s induced by multiple similarities.

\section{Multiple-similarities-induced exceptional structures}
\label{s5}

Imposing multiple similarities simultaneously on the system reduces the codimension of exceptional structures further than what any individual similarity achieves.
The combinations of any two different similarities yield identical overall spectral symmetries, namely $\{\lambda\}=\{\lambda^*\}$, $\{\lambda\}=\{-\lambda^*\}$, and $\{\lambda\}=\{-\lambda\}$.
The consequences of this highly symmetric situation are presented in this section.

\subsection{Implication of triple similarity and likeness of  \texorpdfstring{$\bm{2n}$}{2n}- and  \texorpdfstring{$\bm{(2n+1)}$}{2n+1}-band models}

An important observation is that any possible pair of two similarities always implies the presence of the third as any combination of two of the spectral symmetries results in the third.
Hence, every system with multiple similarities is at least locally self-skew-similar everywhere, and we again find the likeness of $2n$- and $(2n+1)$-band systems with the $(2n+1)$th band being completely flat.
The degeneracy of all exceptional structures at $\lambda=0$ found in $2n$-band systems subject to multiple similarities is increased by one upon adding the $(2n+1)$th band, while exceptional structures away from the origin remain unchanged.
This allows us to derive all exceptional structures in $2n$- and $(2n+1)$-band systems simultaneously.

The necessary presence of self-skew-similarity further allows us to reduce the degree of the characteristic polynomial of a model subject to multiple similarities.
A $2n$-band model $H_{2n}$ hosting multiple similarities has a characteristic polynomial of the form
\begin{equation} \label{eq:CharPolMS}
    P_{2n}(\lambda) = \left(\lambda^{2n} - \sum_{j=0}^{n-1} a_{2j} \lambda^{2j}\right),
\end{equation}
with $a_{2j}$ being real-valued functions of some parameter space. 
Using the substitution
\begin{equation}
    z = \lambda^2 - \frac{a_{2n-2}}{n}
\end{equation}
the characteristic polynomial can be transformed to
\begin{equation}\label{eq:red_pol_MS}
    \Tilde{P}_n(z) = z^n - \sum_{j=0}^{n-2} b_j z^j \, ,
\end{equation}
with $b_j = b_j(a_0,\dots,a_{2n-2})\in \mathbb{R}$.
Given the $n$ solutions $z_l$, the eigenvalue of $H_{2n}$ takes the form
\begin{equation}
    \lambda_\pm(z_l) = \pm \sqrt{z_l +a_{2n-2}/n}.
\end{equation}
Consequently, multiple-similarities-induced EP$2n$s are of codimension $n$.

\subsection{Exceptional structures induced by multiple similarities in three dimensions}

Using the method above, we will now investigate what exceptional structures are induced by multiple similarities in three dimensions, thus turning to a six-band model $H$ and define
\begin{align}
    \kappa &= -\frac{\tr(H^2)}{6} \, , \\
    \eta &= \frac{1}{72}\left\{ \left[\tr(H^2)\right]^2-6\tr(H^4)\right\} \, , \\
    \nu &= -\frac{5\left[\tr(H^2)\right]^3}{864} + \frac{\tr(H^2)\tr(H^4)}{48} - \frac{\det(H)}{2} \, .
\end{align}
The similarities lead to $\kappa,\eta,$ and $\nu\in\mathbb{R}$ in terms of which the characteristic polynomial is given by
\begin{equation}
    \mathcal{P}_6 (\lambda) = \lambda^6+3\kappa \lambda^4 + 3(\eta+\kappa^2) \lambda^2 +\left(\kappa^3+3\kappa\eta- 2\nu\right) \, .
\end{equation}
The discriminant reads
\begin{equation}
    \mathcal{D}_6 = -746496 \left(\eta^3+\nu^2\right)^2\left(\kappa^3+3\eta\kappa-2\nu\right) \, .
\end{equation}
Employing the previously introduced mapping $\lambda=\pm\sqrt{z-\kappa}$ results in the reduced real polynomial
\begin{equation}
    \tilde{\mathcal{P}}_6(z) = z^3 +3\eta z - 2\nu \, .
\end{equation}
Solving $\tilde{\mathcal{P}}_6=0$ for $z$,  the exceptional structure of $H$ can be derived. 
With $\alpha_\pm = (\nu \pm \sqrt{\eta^3+\nu^2})^{1/3}$, $\beta=(1+i\sqrt{3})/2=\exp(i\pi/3)$, and
\begin{equation}
    \begin{split}
        z_1 &= \alpha_+ + \alpha_-, \quad
        z_2 = -\beta^* \alpha_+ - \beta \alpha_-, \\
        z_3 &= -\beta \alpha_+ - \beta^* \alpha_- \, ,
    \end{split}
\end{equation}
the six eigenvalues are given by
\begin{equation}
    \lambda_{j,\pm} = \pm\sqrt{z_j-\kappa} \, ,
\end{equation}
where $j=1,2,$ and $3$.

\begin{figure*}
\includegraphics[width=\textwidth]{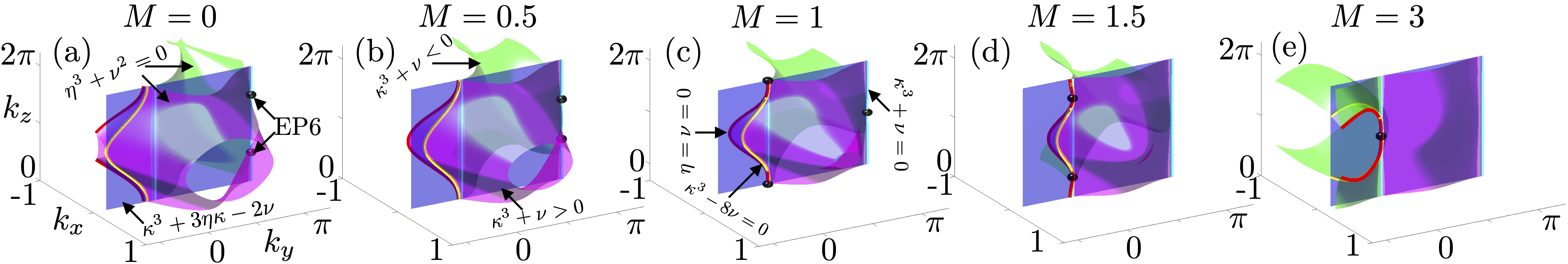}
\caption{Exceptional structures in part of the Brillouin zone of the model given by Eqs.~\eqref{eq:MSHam1}-\eqref{eq:MSHam3}. As predicted, the EP6s (black dots) are connected by various exceptional structures. The blue surfaces denote EP2s corresponding to $\kappa^3+3\eta\kappa-2\nu=0$, while the green (purple) surfaces are EP2s corresponding to $\eta^3+\nu^2=$ for $\kappa^3+\nu<0$ ($\kappa^3+\nu>0)$. 
On these surfaces are special EP2 arcs, one of which is located at $\lambda=0$, defined by the additional constraint $\kappa^3-8\nu$ (yellow arcs). Along these arcs, all eigenvalues are twofold degenerate. Additionally, if instead $\kappa^3+\nu=0$ on the EP2 surfaces, arcs of EP4s at $\lambda=0$ appear (cyan arcs). When $\eta = \nu = 0$, arcs of either purely real or complex conjugated EP3s emerge (red arcs). All of these structures are possibly connected to EP6s, but as the figure shows, they are not always. For $M=0,0.5$ [panels (a) and (b), respectively], the EP4 arcs connect the EP6s, while the EP3 and special EP2 arcs do not. When this EP6 pair is annihilated at $M=1$ [panel (c)], another pair is created. These EP6s are connected by all different EP arcs; see panel (d) for $M=1.5$. At $M=3$, these EP6s are annihilated, resulting in the EP arcs to disconnect. 
\label{fig:MSEP} }
\end{figure*}

The similarity-induced EP6s emerge as pairs if the constraints $\kappa=\eta=\nu=0$ are met. 
There are multiple features in the exceptional structure connecting an EP6 pair:
\begin{itemize}
    \item[(1)] Arcs along which pairs of EP3s emerge, which are either purely real or imaginary;
    \item[(2)] surfaces with pairs of real or imaginary EP2s;
    \item[(3)] arcs of EP4s at $\lambda=0$, which mark one type of intersection between the different EP2 surfaces;
    \item[(4)] arcs on which there are three distinct eigenvalues, which are all twofold degenerate. These arcs appear on all intersections of different EP2 surfaces, which do not result in EP4 arcs;
    \item[(5)] surfaces with a single isolated EP2 at $\lambda=0$.
\end{itemize}
No other exceptional structures can emerge in six-band systems in the presence of multiple similarities. 
Although EP5s have codimension 2 in presence of multiple similarities (cf. Table~\ref{tab:similarities}) and thus are expected to appear as 1D arcs in a 3D parameter space, the spectral symmetry prevents the EP5s from emerging in a six-band system, (cf. Table~\ref{tab:sim_def}). 

Let us explain how these structures arise in more detail in the following.
First, we consider $\eta=0$ and $\nu=0$ with $\kappa\neq0$.
These constraints define two arcs in the 3D parameter space on which EP3s emerge (1). 
The arcs connect the EP6 pair and the sign of $\kappa$ determines whether the EP3s are real or imaginary.
Similarly to the isolated EP2 arcs in pseudo-Hermitian four-band systems [cf. Sec.~\ref{s3} and the yellow arcs in Figs.~\ref{fig:EPPSH}(a)-(e)], the EP3 arcs here have a codimension different than expected, the reason being that the EP3s are not induced by both similarities. If the EP3s  are real they are induced by the spectral constraint $\{\lambda\}=\{\lambda^*\}$ while the second spectral constraint $\{\lambda\}=\{-\lambda^*\}$ or $\{\lambda\}=\{-\lambda\}$ results in the symmetric appearance of the EP3 arcs.
The single spectral constraint reduces the codimension only to 2, and not to 1, and thus the EP3s appear as arcs.
For the imaginary EP3s, $\{\lambda\}=\{-\lambda^*\}$ induces the EP3s and the pairwise appearance fulfills the other constraints, which results in codimension 2 for these EP3s as well.

If $\eta^3+\nu^2=0$ with $\eta\neq0\neq\nu$, we find  pairs of EP2 surfaces (2).
For $-\nu>\kappa^3$, the EP2 pairs are on the real axis, and we find an imaginary pair of EP2s if $-\nu<\kappa^3$. 
In general, the remaining eigenvalues are nondegenerate, real for $8\nu>\kappa$, and imaginary for $8\nu<\kappa$.

There are two additional constraints we have to take into account for the $\eta^3+\nu^2=0$ surface: $\kappa^3-8\nu=0$ and $\kappa^3+\nu=0$.
If $\kappa^3-8\nu=0$, there is an additional EP2 at $\lambda=0$, where $z_i-\kappa=0$. 
This defines an arc in the 3D parameter space on which each eigenvalue is twofold degenerate (3). 
When instead $\kappa^3+\nu=0$, the surface pair of EP2s merge into a single EP4 at the origin of the complex eigenvalue plane (4).
This arc connects pairs of EP6s and marks the seam between the surface with real and imaginary EP2 pairs.

If the second factor in the discriminant vanishes, $\kappa^3+3\eta\kappa-2\nu=0$, we always find a degeneracy at $\lambda=0$ fulfilling $z_i-\kappa=0$ for one of the values of $z_i$.
This defines another surface of EP2s on which the remaining eigenvalues can be real, imaginary, or complex (5).
If $\eta^3+\nu^2\neq0$, the remaining eigenvalues must be nondegenerate.
If the $\kappa^3+3\eta\kappa-2\nu=0$ surface intersects with the $\eta^3+\nu^2=0$ surface, we find the $\kappa^3-8\nu=0$ and $\kappa^3+\nu=0$ arcs discussed before.

Overall, we find that similarity-induced EP6 pairs are connected by multiple EP2 surfaces, on which there are two EP3 arcs and two EP4 arcs. 
Six-band systems with multiple similarities do not host any further exceptional structures.
There are, however, notable Fermi structures discussed in SM2 in the Supplemental Material~\cite{Supmat}.
Due to the high spectral symmetry, the number of nontrivial Fermi structures is low for multiple similarity constrained models.

\subsection{An example}
To illustrate the exceptional structures induced by multiple similarities in 3D, we study the model given by
\begin{align}
    H_6&=\begin{pmatrix} 0&1&0&0&0&0\\0&0&1&0&0&0\\0&0&0&1&0&0\\0&0&0&0&1&0\\0&0&0&0&0&1\\a_0&0&a_2&0&a_4&0\end{pmatrix} \, ,\label{eq:MSHam1}
    \\
    a_0 &= \sin(k_x), \quad a_2 = \sin(k_y), \label{eq:MSHam2}
    \\
    a_4 &= M-\cos(k_x)-\cos(k_y)+\cos(k_z), \label{eq:MSHam3}
\end{align}
with $k_x,k_y,k_z$ the lattice momentum components and $M\in \mathbb{R}$.
The corresponding exceptional structures are displayed in Fig.~\ref{fig:MSEP} and confirm the aforementioned general statements.
The pairwise creation/annihilation of the EP6s as a function of the parameter $M$, which is swept through in Figs.~\ref{fig:MSEP}(a)-(e), suggests that these are topological, a topic that we discuss in the following subsection.

Finally, we comment on how these exceptional structures are affected upon adding a seventh band, which due to the similarities necessarily is flat and appears at $\lambda=0$.
As mentioned earlier, this promotes all degeneracies appearing at $\lambda=0$ by one, while the degeneracies of those appearing at finite $\lambda$ remain the same.
Thus, the degeneracies of the surfaces of EP2s and the arcs of EP3s remain the same at $\lambda\neq0$, while the EP6s, the arcs of EP4s, and the surfaces of EP2s at $\lambda=0$ are promoted to EP7s, EP5s, and EP3s, respectively. 
The special arc, along which all eigenvalues are twofold degenerate, illustrated by a yellow arc in Fig.~\ref{fig:MSEP}, keeps the pair of nonzero twofold degenerate eigenvalues, and the eigenvalue at $\lambda=0$ is promoted to a threefold degeneracy. 

\subsection{Dimensional generalization}
\begin{figure}
\includegraphics[width=0.9\columnwidth]{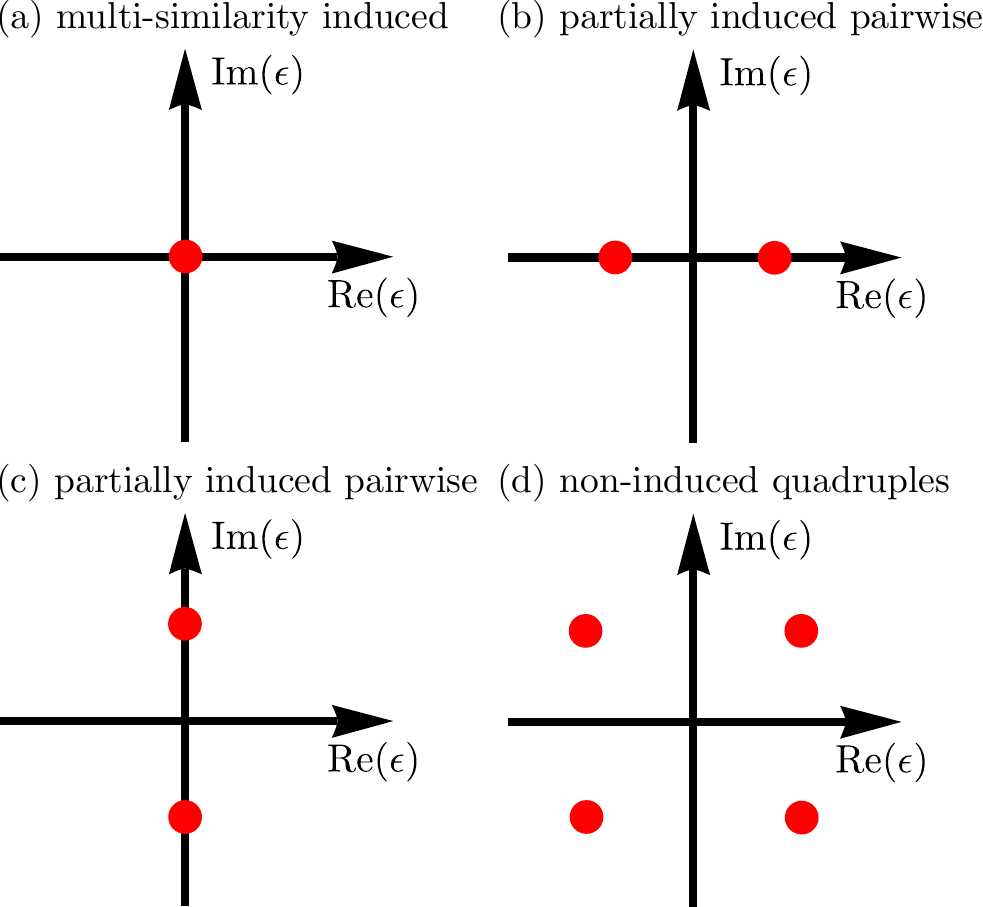}
\caption{Illustration of how multiple similarities constrain the position of exceptional structures of $n$-band systems (red dots) in the complex eigenvalue plane. (a) EP$m$s induced by multiple similarities are forced to appear at the origin and have codimension $k$ when $m=2k+\delta$ [$\delta=0(1)$ for $m\in\text{even}(\text{odd})$]. When the EP$m$ is induced only by one similarity relation, its location in the complex eigenvalue plane is bound to satisfy the spectral relation following the remaining similarity. EP$m$s of codimension $m-1$ induced only by pseudo-Hermiticity [panel (b)] or self-skew-similarity [panel (c)] can emerge when $m<\frac{n}{2}$, and are bound to appear at purely real or imaginary eigenvalues of opposite signs, respectively. (d) When $m<\frac{n}{4}$, EP$m$s of generic codimension $2m-2$ can emerge, and these necessarily appear in all four quadrants.
\label{fig:Loc_MS} }
\end{figure}

Just as for pseudo-Hermiticity and self-skew-similarity, systems subject to multiple similarities not only host the exceptional structures induced by multiple similarities, but other structures emerge in the spectrum as well.
To illustrate how these emerge, we resort to similar methods as used in previous sections, keeping in mind that this case is slightly more subtle. 
The spectral constraints imposed by multiple similarities are fulfilled by the following three different operators:
\begin{align}
    H_1 &= \begin{pmatrix} \mathcal{H}_{\text{PsH}} &0\\0&-\mathcal{H}_{\text{PsH}} \end{pmatrix}\, , \label{eq:PsHSSS}
    \\
    H_2 &= \begin{pmatrix} \mathcal{H}_{\text{SSS}}&0\\ 0& \mathcal{H}_{\text{SSS}}^{\dagger} \end{pmatrix} \, , \label{eq:SSSPsH}
    \\
    H_3 &= \begin{pmatrix} \mathcal{H} &0&0&0\\0&\mathcal{H}^{\dagger}&0&0\\0&0&-\mathcal{H} &0\\ 0&0&0&-\mathcal{H}^{\dagger} \end{pmatrix} \, , \label{eq:generic}
\end{align}
with $\mathcal{H}_{\text{PsH}}$ and $\mathcal{H}_{\text{SSS}}$ being pseudo-Hermitian and self-skew-similar $n/2$-band models, respectively, and $\mathcal{H}$ is a generic non-Hermitian operator with $n/4$ bands.
Hence, $H_1$, $H_2$, and $H_3$ describe $n$-band models.
The form of $H_1$ indicates that the multiple similarity-induced exceptional structures are accompanied by codimension $m-1$ EP$m$s induced by pseudo-Hermiticity only, while the form of $H_2$ further suggests the existence of self-skew-similarity induced EP$m$s of codimension $m$, where $m\leq n/2$ in both cases.
Because of the self-skew-similarity, the pseudo-Hermiticity-induced EP$m$s necessarily appear in pairs in adjacent quadrants of the complex eigenvalue plane, while the pseudo-Hermiticity enforces the self-skew-similarity-induced EP$m$s to appear at purely real eigenvalues, or in pairs at complex conjugate eigenvalues.
Finally, the form of $H_3$ shows that operators subject to multiple similarities  may also host generic EP$m$s of codimension $2m-2$.
Although the form of $H_3$ is, in principle, covered as a special case of $H_1$ and $H_2$, writing it out explicitly makes it clear that the EPs of generic codimension display a double mirror symmetry in the complex eigenvalue plane, as they necessarily appear in all four quadrants of the complex plane.
The one exception to this is when the exceptional structures appear at the real (imaginary) eigenvalue axis, resulting in a pairwise appearance at some eigenvalue $\lambda = \pm E_0 \in \mathbb{R}\; (\in i\mathbb{R})$.
This is because of the respective spectral constraint; pseudo-Hermiticity (pseudo-anti-Hermiticity) induces a symmetry with respect to the real (imaginary) axis, while self-skew-similarity induces a symmetry with respect to the arcs $\text{Re}(\lambda) = + (-) \text{Im}(\lambda)$ if the signs of the real and imaginary parts of $\lambda$ are equal (opposite).
How the emergence of the different exceptional structures is constrained in the complex eigenvalue plane is displayed in Fig.~\ref{fig:Loc_MS}.

Upon adding an additional band to $H_1$, $H_2$, or $H_3$, this is by the similarities enforced to be flat and located at $\lambda=0$.
Thus, all exceptional degeneracies appearing at zero eigenvalue will be increased by one.
The addition of two bands for $H_1$ or $H_2$ increases the dimension of $\mathcal{H}$ by one.
As for $H_3$, the case is slightly more subtle, as adding two additional bands reduces the model to either $H_1$ or $H_2$.
Only by adding four additional bands, the structure of $H_3$ can be preserved.

This comprises all possible combinations of exceptional structures in systems subject to multiple similarities.
To complement the 3D example of the previous subsection, we in SM3 in the Supplemental Material~\cite{Supmat} present the technical details  of the 4D case.

\subsection{Topological classification of multiple-similarity-induced EP\texorpdfstring{$\bm{n}$}{n}s in  \texorpdfstring{$\bm{n}$}{n}-band systems}
As hinted by the above six-band example in 3D, where the EP6s come in pairs, EP$n$s protected by multiple similarities emerging in $n$-band models are \emph{topological}.
By extending the classification schemes recently developed in Refs.~\cite{Yoshida2024,Stalhammar2024}, we show that the Abelian eigenvalue topology of EP$n$s is captured by the resultant winding number.
We also map out the vector bundle classification and identify the source to the eigenvalue topology in terms of Clifford bundles.

\subsubsection{Resultant vector and winding number}

The resultant winding number is the winding of the resultant vector, in which different resultants of the characteristic polynomial and its corresponding derivatives are collected. 
For a system subject to multiple similarities, the resultant components from a characteristic polynomial of the form in Eq.~\eqref{eq:CharPolMS} are defined as
\begin{equation}
    r_j = \text{Res}\left[\partial^{n-2j}_{\lambda}P_n(\lambda),\partial^{n-1}_{\lambda}P_n(\lambda)\right],
\end{equation}
with $j\in \{1,\dots,\lfloor \frac{n}{2}\rfloor\}$, where $\lfloor x\rfloor$ is the floor function and hence  denotes the integer part of $x$.
The resultant vector then reads
\begin{equation}
    \mathbf{R} = \left(r_1,...,r_{\lfloor \frac{n}{2}\rfloor}\right).
\end{equation}
Importantly, the resultant vector $\mathbf{R}$ is constructed in such a way that it vanishes exactly at (and only at) the points in momentum space corresponding to EP$n$s in the parent non-Hermitian Hamiltonian~\cite{Yoshida2024,Stalhammar2024}.
Therefore, the winding number of the resultant vector around these points can be thought of as ``monopole charges'' of the EP$n$s. 
The winding number corresponds to the degree of a map between spheres in the parameter space and the space of resultants,
\begin{align}
S^{\lfloor \frac{n}{2}\rfloor-1} &\to S^{\lfloor \frac{n}{2}\rfloor-1},\\
    \frac{\bm k}{| \bm k |}&\mapsto \frac{\bm R}{| \bm R |},
\end{align}
and counts exactly how many times the resultant vector winds around the EP$n$.
As such, this is a topological invariant given by $\pi_{\lfloor \frac{n}{2}\rfloor-1}\left(S^{\lfloor \frac{n}{2}\rfloor-1}\right) = \mathbb{Z}$, when $n>3$, and thus we expect an integer-valued topological invariant classifying these EP$n$s.
Up to a normalization factor, the winding number is given by
\begin{equation}
    W \propto \oint_{S^{\lfloor \frac{n}{2}\rfloor-1}} \text{Tr}\left[\left(\mathbf{n}^{-1}d\mathbf{n}\right)^{\lfloor \frac{n}{2}\rfloor-1}\right],
\end{equation}
with $\mathbf{n} = \frac{\mathbf{R}}{|\mathbf{R}|}$ the normalized resultant vector.
From this winding number follows also a doubling theorem, which can be shown as (using Stokes theorem)
\begin{align}
    \sum_k W_k &\propto \sum_k \oint_{S_k^{\lfloor \frac{n}{2}\rfloor-1}} \text{Tr}\left[\left(\mathbf{n}^{-1}d\mathbf{n}\right)^{\lfloor \frac{n}{2}\rfloor-1}\right] \nonumber
    \\
    &= \int_{\text{BZ}\setminus \Delta}d \text{Tr}\left\{\left[d\ln(\mathbf{n})\right]^{\lfloor \frac{n}{2}\rfloor-1}\right\}=0,
\end{align}
where the last step holds using the second Bianchi identity. 
Here, $\text{BZ}\setminus \Delta$ denotes the punctured Brillouin zone, i.e., the Brillouin zone with the EP$n$s removed.

Although the above reasoning is in principle applicable also when $n=2,3$, it is worth commenting on these cases separately.
Since $\pi_0(S^0)=\mathbb{Z}_2$, we no longer expect that an integer invariant classifies two and threefold EPs induced by multiple similarities.
Very much like EP2s induced by pseudo-Hermiticity only (see Ref.~\cite{Stalhammar2024} for a detailed description), these are instead classified by a $\mathbb{Z}_2$ invariant, the origin of which is the introduction of integration on $S^0$.
Since $S^0$ comprises a set of two points, integration is defined as a weighted sum of the integrand evaluated at these points.
Therefore, the corresponding winding number will take the values $\pm 1$ always, and the doubling theorem still holds.

\subsubsection{Resultant Hamiltonian and vector bundle classification}
Finally, we unravel the topological nature and origin of the resultant winding number classifying EP$n$s protected by multiple similarities.
The key here lies in the fact that the resultant vector can be used to construct a Hermitian Hamiltonian, known as the resultant Hamiltonian~\cite{Stalhammar2024},
\begin{equation}
    H_{\mathbf{R}}(\bk) = \sum_{j=1}^{\lfloor\frac{n}{2}\rfloor} r_j \gamma ^j,
\end{equation}
with $\gamma^j$ being Clifford algebra generators. 
By construction, this Hamiltonian hosts nodal points exactly at the same points in momentum space as where the parent non-Hermitian system hosts EP$n$. As such, the resultant Hamiltonian allows for an interpretation of the non-Hermitian eigenvalue topology in terms of the Hermitian eigenvector topology in the context of the Altland-Zirnbauer symmetry classes~\cite{Altland1997,Ryu2010}. 
It should be noted that since the spectrum of $H_{\mathbf{R}}(\bk)$ is given by $\pm | \mathbf{R}(\bk)|$, the resultant Hamiltonian displays a ``generalized spin degeneracy'', as is customary for operators of that form~\cite{Mathai2016}.

Which symmetry group determines the topology of these nodal points (and consequently also the corresponding EP$n$s) is different for different values of $n$.
When $\lfloor\frac{n}{2}\rfloor=2m-1$ for some integer $m$, i.e., when it is odd, $H_{\mathbf{R}}(\bk)$ is expressed as a linear combination of all generators of the Clifford algebra of dimension $2^{\frac{1}{2}\left(\lfloor \frac{n}{2}\rfloor+1\right)}\times 2^{\frac{1}{2}\left(\lfloor \frac{n}{2}\rfloor+1\right)}$.
Hence, $H_{\mathbf{R}}(\bk)$ falls within the symmetry class A of the Altland-Zirnbauer classification. 
The nodal points appear in odd dimensions (since their codimension $\lfloor\frac{n}{2}\rfloor$ is odd), and are therefore expected to be classified by even-dimensional topological invariants stemming from symmetry class A.
These are known to take integer values, and correspond to Chern numbers.

When $\lfloor\frac{n}{2}\rfloor$ is even, the resultant Hamiltonian lacks one of the $\gamma$ matrices in its definition.
The remaining $\gamma$ matrix can instead be used to define a chiral symmetry, since 
\begin{equation}
    \left\{H_{\mathbf{R}}(\bk),\gamma^{\lfloor \frac{n}{2}\rfloor +1}\right\}=0.
\end{equation}
Along the same line of reasoning, the nodal points will in this case be classified by odd-dimensional topological invariants stemming from the AIII symmetry class, which are integer-valued generalized winding numbers.

When $n=2,3$, the situation is completely different.
The resultant Hamiltonian is then a scalar and does not satisfy chiral symmetry.
Most notable is perhaps that EP2s and EP3s induced by multiple similarities thus are not related to Hermitian nodal points through the resultant Hamiltonian, but rather to the zeros of a one-band Hermitian Hamiltonian.
This scalar will, however, display other important features when evaluated on the ``sphere'' surrounding these zeros, since the zero-sphere only comprises two points.
Evaluating $H_{\mathbf{R}}(\bk)$ on $S^0$ is done as
\begin{equation}
    H_{\mathbf{R}}(k)_{|_{S^0}} = \begin{cases}H_{\mathbf{R}}(-p)\sigma(-p)&\\H_{\mathbf{R}}(p)\sigma(p)& \end{cases} = \begin{cases} -H_{\mathbf{R}}(-p)&\\H_{\mathbf{R}}(p)&\end{cases}.
\end{equation}
If we now let $\bk\to -\bk$, $H_{\mathbf{R}}(\bk)\xrightarrow[]{\bk\to -\bk} -H_{\mathbf{R}}(-\bk)$.
Thus, the resultant Hamiltonian in this case satisfies a particle-hole symmetry with identity as generator, and therefore falls into symmetry class D.
This agrees with the previous observation that the resultant winding number only attains values of $\pm 1$. The invariant stemming from symmetry class D is indeed known to be a $\mathbb{Z}_2$ invariant.

As a final, somewhat technical, comment regarding the topological origin of the resultant winding number, we also state the vector bundle classification of EP$n$s induced by multiple similarities.
In fact, these can be read off directly from the codimension of the EP$n$, as listed in Ref.~\cite{Stalhammar2024}.
In general, the correct vector bundle for classifying EP$n$s is the tangent bundle of $\mathbb{T}^{\lfloor \frac{n}{2}\rfloor}$, i.e., $T\mathbb{T}^{\lfloor \frac{n}{2}\rfloor}$, meaning that the resultant vector can be thought of as a vector field on $\mathbb{T}^{\lfloor \frac{n}{2}\rfloor}$, since $\mathbf{R}$ is a section of the tangent bundle denoted by $\mathbf{R} \in \Gamma \left( T \mathbb{T}^{\lfloor\frac{n}{2}\rfloor}\right)$.
Thus, the resultant vector for EP2s and EP3s is a vector field on the circle $S^1$, for EP4s and EP5s it is a vector field on the torus $\mathbb{T}^2$, and so on.
For technical details behind the vector bundle construction, we refer to Refs.~\cite{Stalhammar2024,Mathai2016}.
The contents of this section are summarized in Table~\ref{tab:sumres}.

\begin{table}
\begin{tabular}{c | c | c | c}
\hline
\hline
Bands &$n=2,3$& {$\lfloor \frac{n}{2}\rfloor$ even} & $\lfloor \frac{n}{2}\rfloor$ odd\\
 \hline
 \hline
 Inv. & $\mathbb{Z}_2$ & $\mathbb{Z}$ & $\mathbb{Z}$  \\
 \hline
  
   \begin{tabular}{@{}l@{}} 
   \\
   Res. \\
   Ham.
   \\
   \\
   \end{tabular}
   &
  \begin{tabular}{@{}l@{}} 
   $H_{\mathbf{R}} = R$
   \end{tabular}& 
   \begin{tabular}{@{}l@{}} 
   $H_{\mathbf{R}} = \sum_{i=1}^{\lfloor \frac{n}{2}\rfloor}r_i \gamma^i$, \\
   $\left\{H_{\mathbf{R}},\gamma^{\lfloor\frac{n}{2}\rfloor +1 }\right\} = 0$
   \end{tabular} & 
   \begin{tabular}{@{}l@{}} 
   $H_{\mathbf{R}} = \sum_{i=1}^{\lfloor \frac{n}{2}\rfloor}r_i \gamma^i$
   \end{tabular} 
   \\
  \hline
  \begin{tabular}{@{}l@{}}
  Herm.
  \\
    Deg.
  \end{tabular}
  & $2^0$ & $2^{\frac{1}{2}\left(\lfloor \frac{n}{2}\rfloor+1\right)}$ & $2^{\frac{1}{2}\left(\lfloor \frac{n}{2}\rfloor+1\right)}$ \\
  \hline
   \begin{tabular}{@{}l@{}}
    Sym. 
    \\
    class
    \end{tabular}
    & D & AIII & A \\
   \hline
   \begin{tabular}{@{}l@{}} 
    Vector 
    \\
    bundle
 \end{tabular}
   & $TS^1$, rank 1 &  $T \mathbb{T}^{\lfloor \frac{n}{2}\rfloor}$, rank $\lfloor \frac{n}{2}\rfloor$ & $T \mathbb{T}^{\lfloor \frac{n}{2}\rfloor}$, rank $\lfloor \frac{n}{2}\rfloor$
   \\
   \hline
    \begin{tabular}{@{}l@{}}
   Vector 
   \\
   field
\end{tabular}
   & $R \in \Gamma\left( TS^1\right)$ & $\mathbf{R} \in \Gamma\left( T\mathbb{T}^{\lfloor \frac{n}{2}\rfloor}\right)$ & $\mathbf{R} \in \Gamma\left( T\mathbb{T}^{\lfloor \frac{n}{2}\rfloor)}\right)$  \\

   \hline
   \hline
\end{tabular}
 \caption{Summary of the topological nature and origin of the invariants classifying EP$n$s.} \label{tab:sumres}
\end{table}

\section{Concluding Remarks}
\label{s6}
\subsection{Summary}
In this work, we have mapped out the exceptional structures induced by one and multiple generalized similarities in three and four dimensions.
We find that the EP4s induced by pseudo-Hermiticity in three dimensions emerge as special points on similarity-induced arcs of EP3s, which in turn are special arcs on similarity-induced surfaces of EP2s.
In addition, we find special arcs of EP2s, where the spectral constraints manifest as forcing these to emerge either at purely real eigenvalues, or as pairs of complex conjugate eigenvalues, leading to their differing codimension.
Similarly, we find that the EP4s induced by self-skew-similarity in four dimensions are accompanied by different surfaces of EP2s, one located at zero eigenvalue and one located at a finite eigenvalue.
Consequently, these behave differently upon adding a fifth band at zero eigenvalue, promoting the EP4s to EP5s, but only one of the EP2 surfaces to a surface of EP3s.

When exposed to multiple similarities, a system displays an even more exotic plethora of exceptional structures.
In three dimensions, the similarity-induced EP6s are found to emerge as special points on similarity-induced arcs of EP4s at zero eigenvalue, which in turn emerge as special arcs on similarity-induced surfaces of EP2s.
Additionally, these are accompanied by exceptional structures that are induced by only one of the similarities, giving paired arcs of EP3s emerging at purely real or imaginary eigenvalue, arcs of EP2s where three distinct eigenvalues are twofold degenerate, and a single surface of EP2s at zero eigenvalue.
Notably, the spectral constraints do not allow EP5s to appear in this setting.
When adding a seventh flat band, the exceptional degeneracies appearing at zero eigenvalue are increased by one.

Our specific studies also allow us to make statements in an arbitrary number of dimensions.
We are led to conclude that pseudo-Hermitian systems, in addition to similarity-induced codimension $k-1$ manifolds of EP$k$s, also host generic codimension $2(l-1)$ manifolds of EP$l$s, on which the spectral symmetries are reflected as enforcing these to either real eigenvalues, or that they emerge as pairs of complex conjugate eigenvalues (cf. Fig.~\ref{fig:Loc_PsH}).
Similarly, in self-skew-similar systems, the similarity-induced EP$k$s of codimension $\lfloor k \rfloor$ are accompanied by generic codimension $2(l-1)$ manifolds of EP$l$s, where the spectral symmetries determine the location in the complex eigenvalue planes of the EP$l$s (cf. Fig.~\ref{fig:Loc_SSS}).
Along an identical line of reasoning, the codimension $\lfloor \frac{k}{2} \rfloor$ EP$k$s induced by multiple similarities are accompanied by exceptional structures of EP$ls$ induced by pseudo-Hermiticity (codimension $l-1$) and self-skew-similarity (codimension $\lfloor l \rfloor$) alone, and generic EP$j$s (codimension $2(j-1)$), where the combined similarities constrain their emergence in the complex eigenvalue plane, as illustrated in Fig.~\ref{fig:Loc_MS}.
Finally, we provide a classification of the Abelian eigenvalue topology of EP$n$s induced by multiple similarities in $n$-band systems using the resultant winding number.

\subsection{Discussion and outlook}
Apart from providing a missing piece of forefront theoretical research within non-Hermitian topological band theory and providing a complete insight of the emergence of similarity-induced exceptional structures in arbitrary dimensions, our work is of experimental relevance in several modern physical platforms.
Starting with the cases of single similarities, $\mathcal{PT}$-symmetric systems are an important special case following the studies of pseudo-Hermitian similarity~\cite{Montag2024_2}.
In optics, $\mathcal{PT}$ symmetry manifests as a perfect balance between gain and loss in photonic crystals~\cite{ozdemir2019,Lu2014,Ozawa2019}, where our work can be used to increase the fundamental understanding of systems with three or more variable system parameters.
Moreover, $\mathcal{PT}$-symmetric systems can also be constructed using electrical circuits, where the corresponding Laplacian reflects the symmetry if resistivity is introduced in a balanced way~\cite{gupta2021}. 
Thus, topolectrical circuits comprise a suggestive class of systems, where parts of our theoretical predictions could be experimentally probed.
Other promising candidates include single-photon interferometry, where symmetry-protected rings of EP2s~\cite{Wang2021}, and EP3s~\cite{Wang2023} have been demonstrated, spin-orbit-coupled cold atoms, where EP3s are found~\cite{Liu2024}, coupled acoustic cavities hosting arcs of EP3s on EP2 surfaces~\cite{Tang2023}, nitrogen-vacancy spin systems, so far realizing arcs of EP3s~\cite{Wu2024}, and correlated quantum many-body systems, where EP4s induced by symmetry-preserving interactions have been theoretically predicted in three spatial dimensions~\cite{Schafer2022}. 
Similarly, anti-$\mathcal{PT}$ symmetric systems are the special case of pseudo-anti-Hermitian systems that have received most experimental attention so far.
Recently, EP3s have been studied in anti-$\mathcal{PT}$ symmetric optical microcavities~\cite{Jahangiri2025}. 
Further, in both optical parametric systems~\cite{Li2025} and stimulated Brillouin scattering~\cite{Bergman2021} EP2s have been induced by anti-$\mathcal{PT}$ symmetry. 
An open question is how to employ our results to induce higher-order EPs in these platforms.
Moving to self-skew-similarity, it comprises a generalization of the physically important sublattice symmetry, commonly present in non-Hermitian Lieb lattices~\cite{Vicencio2015,Mukherjee2015,Xiao2020}, yet another important non-Hermitian extension of optics. 

Concerning systems subject to multiple similarities, a very recent work suggests that Kerr ring resonators serve as a viable experimental platform~\cite{Hill2025}.
Although present progress is limited to a 2D parameter space, a 3D generalization of such a setup is a good candidate to realize the features predicted in Sec.~\ref{s5} well within experimental reach.

Apart from experimental relevance and potential future experimental advances, there are additional theoretical directions to investigate further.
One such example is to combine the local similarities studied in this work, with nonlocal versions, including non-Hermitian time-reversal symmetries, particle-hole symmetries, and inversion symmetry~\cite{Sayyad2022}.
The Abelian eigenvalue topology of EP$n$s induced by these symmetries was recently classified in terms of the Hermitian tenfold way~\cite{Altland1997,Schnyder2008,Kitaev2009,Ryu2010}, utilizing a mapping between non-Hermitian eigenvalue topology and Hermitian eigenvector topology~\cite{Stalhammar2024}.
Although not reducing the codimension of the EPs~\cite{Sayyad2022}, these symmetries drastically change the topological properties of EP$n$s in $n$-band systems when no other similarities/symmetries are present~\cite{Stalhammar2024}, and suggestively they should do the same for the less degenerate exceptional structures on which these EP$n$s reside.
Also, it is not understood whether these symmetries further constrain EP$m$s, $m<n$, to only emerge in certain regions of the complex eigenvalue plane, which opens up possibilities for further studies along these lines.

Conclusively, our work not only provides a missing piece of modern research within theoretical non-Hermitian physics, but also displays how abstract mathematical notions could manifest in realistic laboratory setups thanks to its direct relevance in a large variety of areas in modern experimental physics.
Having a fundamental understanding of the spectral properties provides a powerful and versatile tool for predicting exotic and potentially hitherto unknown phenomena.
Therefore, developing this framework further and deepening the connection between mathematics and physics comprises an important and natural step in expanding the vast subject of topological band theory.

\acknowledgments

M.S. thanks Lukas K\"onig and Lukas R\o dland for stimulating discussions and related collaborations.
A.M. and F.K.K. acknowledge funding from the Max Planck Society Lise Meitner Excellence Program 2.0.
A.M. and F.K.K. also acknowledge funding from the European Union via the ERC Starting Grant “NTopQuant”. 
Views and opinions expressed are, however, those of the authors only and do not necessarily reflect those of the European Union or the European Research Council (ERC). Neither the European Union nor the granting authority can be held responsible for them. M.S. was supported by the Swedish Research Council (VR) under Grant No. 2024-00272. J.I. thanks the Max Planck-uOttawa Centre for Extreme and Quantum Photonics for funding the MPC Internship at the MPI for the Science of Light in Erlangen.

\bibliography{references.bib}

\end{document}